\documentclass[12pt,a4paper]{article}
\pdfoutput=1
\usepackage{jheppub}
\usepackage[utf8]{inputenc}
\usepackage[T1]{fontenc}
\usepackage{comment}

\usepackage{amsmath,physics,float}  
\usepackage{amssymb}  
\usepackage{latexsym}
\usepackage{graphicx,xcolor}

\usepackage{amsfonts}
\usepackage{braket}
\usepackage{mathrsfs}

\usepackage{subfig} 
\usepackage{empheq}

\usepackage[shortlabels]{enumitem}

\usepackage{pgfplots}
\usepackage{natbib}

\usepackage{bigints}

\newcommand{\beq}{\begin{equation}}
\newcommand{\eeq}{\end{equation}} 
\newcommand{\ben}{\begin{eqnarray}}
\newcommand{\een}{\end{eqnarray}}

\newcommand{\ksi}{\ket{\psi}}

\newcommand{\lc}{\left(}
\newcommand{\rc}{\right)}

\usepackage{amsmath}

\usepackage{tikz, pgf}

\newcommand\NN{{\mathcal N}}

\newcommand\ZZZ{{\mathbb Z}}

\newcommand\om{w}

\begin{document}

\title{\boldmath Normalization of D instanton amplitudes in two dimensional 
type 0B string theory}

\author{Joydeep Chakravarty and Ashoke Sen}
\affiliation{International Centre for Theoretical Sciences - TIFR, 
Bengaluru - 560089, India}
\emailAdd{joydeep.chakravarty@icts.res.in, ashoke.sen@icts.res.in}

\begin{abstract}
{We compute the normalization of the D-instanton amplitudes in type 0B string theory in
two dimensions and find perfect agreement with the dual matrix model result.}
\end{abstract} 

\maketitle


\section{Introduction}

Two dimensional type 0B string theory, whose world-sheet theory consists of a
free scalar associated with time direction, its superpartner fermion, a super-Liouville theory
with central charge $27/2$ and the usual b,c, $\beta,\gamma$ ghost system, is expected
to be dual to a matrix theory -- theory of non-relativistic free fermions moving under the
influence of an inverted harmonic oscillator potential, with fermi level reaching the same
height on both sides of the potential\cite{Takayanagi:2003sm,Douglas:2003up}.
Recently, \cite{Balthazar:2022apu} studied D-instanton contribution to the amplitudes 
in this type 0B
string theory, and found perfect agreement with the results in the dual matrix theory except
for one aspect. The overall normalization of the D-instanton induced amplitudes in
string theory is controlled  by the exponential of the annulus 0-point function,
with the boundaries
of the annulus lying on the D-instanton. However the annulus amplitude suffers
from certain infrared divergences in the open
string channel that could not be resolved using the world-sheet formalism.
\cite{Balthazar:2022apu} dealt with this issue by including
an unknown constant $\NN_D$ in the overall normalization of the amplitudes.
$\NN_D$ was then determined by comparing the string
theory results to the results in the free fermion description of the theory. Once this constant
was fixed, there was perfect agreement between the various quantities calculated using the
two descriptions.

In recent years string field theory has proved to be a useful tool in dealing with infrared
divergences in the open string theory on the D-instanton, including in the
annulus zero point function -- see {\it e.g.} 
\cite{Sen:2021tpp,Eniceicu:2022nay,Alexandrov:2022mmy}. In this paper we use
the same approach and get a finite result for the constant $\NN_D$. 
We find perfect agreement between the results obtained this way and the prediction
for $\NN_D$ obtained by
comparison with the results in the dual  matrix model.

\section{Computation of the normalization constant}

We now describe the details of the computation.
Open strings living on the D-instanton of type 0B string theory come from
only the NS sector.
Our starting point is the expression for the exponential of the
annulus partition function on the D-instanton, 
obtained by taking the $R\to\infty$ limit of eq.(3.14) of \cite{Balthazar:2022apu}:
\beq\label{einst}
\exp\left[ -{\int_0^\infty} {dt\over 2t}\right]\, .
\eeq
This has divergence from the $t\to\infty$ limit. As will be discussed shortly, these
can be attributed to propagation of open string zero modes in the Siegel gauge,
and can be dealt with by standard method. This also has divergence from the $t\to 0$
limit associated with the closed string channel. This is a physical infrared divergence that
indicates that the D-instanton induced processes induce non-trivial changes in the vacuum.
Indeed in the language of dual
matrix model, these correspond to the transmission of a single fermion or a hole across
the inverted harmonic oscillator potential, so that a closed string impinged from one side,
represented by a fermion hole pair, does not get reflected as an usual closed
string\cite{DeWolfe:2003qf} .
While this issue may be resolved by expanding the allowed set of asymptotic states,
we shall avoid this problem by working with instanton anti-instanton induced amplitudes,
for which there is no net transfer of fermion number across the potential barrier and hence the
final state can be represented as a collection of usual closed strings. In string
theory the normalization associated with such amplitudes, with the instanton and
the anti-instanton separated by a distance $\Delta x$, is given by\cite{Balthazar:2022apu}:
\beq\label{einsantipre}
Z = \exp\left[ {\int_0^\infty}{dt\over 2t} \left\{ -2 + 2\, e^{2 \pi t  \left({1\over 2} - {1\over 2}
\left({\Delta x)\over 2\pi}\right)^2\right) }\right\}
\right]   \, .
\eeq
Here the $-2$ term inside the curly bracket
is the contribution from the open strings with both ends on the instanton
or both ends on the anti-instanton while the second term is the contribution from open
strings with one end on the instanton and the other end on the anti-instanton.
The divergence from the $t\to 0$ end has now disappeared as expected, but the
divergence from the $t\to\infty$ end remains. 
\cite{Balthazar:2022apu} introduced an arbitrary constant $\NN_D$ to encode the
effect of these divergences and expressed the result as\footnote{The integrations over $x_1$,
$x_2$ are supposed to be done after including the contributions from other world-sheet
components which also depend on $x_1$, $x_2$.}
\beq \label{ezanspre}
Z = \int dx_1 dx_2 {\NN_D^2\over (\Delta x)^2 - (2\pi)^2}\, ,
\eeq
where $x_1,x_2$ are the positions of the instanton and the anti-instanton.
Our goal will be to reinterpret these
divergences in the language of string field theory and use insights from string field
theory to get a finite result for \eqref{einsantipre} and hence determine $\NN_D$.

The results \eqref{einsantipre}, \eqref{ezanspre} were derived in the
$\alpha'=2$ unit\cite{Balthazar:2022apu}. 
We shall use
some results of \cite{Sen:2021tpp} that used $\alpha'=1$ unit.
For this it will be useful to convert
\eqref{einsantipre}, \eqref{ezanspre}
into $\alpha'=1$ unit. This is done by noting that $x$ and $\alpha'$
occur in the combination $x/\sqrt{\alpha'}$. Hence to convert from $\alpha'=2$ unit to
$\alpha'=1$ unit we need to replace $x$ by $\sqrt 2x$. This replaces \eqref{einsantipre},
\eqref{ezanspre} by
\beq\label{einsanti}
Z = \exp\left[ {\int_0^\infty}{dt\over 2t} \left\{ -2 + 2\, e^{2 \pi t  \left({1\over 2} - 
\left({\Delta x)\over 2\pi}\right)^2\right) }\right\}
\right]   \, ,
\eeq
and
\beq \label{ezans}
Z = \int dx_1 dx_2 {\NN_D^2\over (\Delta x)^2 - 2\pi^2}\, ,
\eeq
respectively.

The general strategy for evaluating \eqref{einsanti} will be the same as those used
{\it e.g.} in \cite{Sen:2021tpp,Eniceicu:2022nay,Alexandrov:2022mmy}:
\begin{enumerate}
\item We first use the identities:
\beq\label{eq1}
\int_0^\infty {dt\over 2t} \left( e^{-2\pi h_1 t} - e^{-2\pi h_2 t}\right) = \ln\sqrt{h_2\over h_1}\, ,
\eeq
and 
\beq\label{eq2}
\int {d\phi\over \sqrt {2\pi}} e^{-{1\over 2} h\, \phi^2} = {1\over \sqrt{h}}, \qquad
\int dp dq e^{-h p q} = h\, ,
\eeq
for grassmann even variable $\phi$ and grassmann odd variables $p,q$, to
express \eqref{einsanti} as path integral over open string fields in the Siegel gauge.
The divergence in the $t\to\infty$ limit comes from states with $h=0$. 
\item For open strings with both ends on the D-instanton,
one $h=0$ state comes from the translation mode of the D-instanton and
gives a contribution of 1 to the coefficient of $dt/(2t)$ term in the integrand. We also
have a pair of grassmann odd zero modes from the ghost sector that contribute
$-2$ to the coefficient of $dt/(2t)$ in the integrand, producing a net factor of
$-\int dt/(2t)$ as in \eqref{einst}. These grassmann odd modes are the Faddeev-Popov
ghost fields  associated with Siegel gauge fixing and the vanishing of $h$
indicates that the Siegel gauge choice breaks down for open strings with both ends on the
D-instanton (or both ends on the anti-D-instanton).
\item We remedy this by replacing the Siegel gauge fixed path
integral by the path integral over the classical open string fields (of ghost number 1)
weighted by the exponential of the gauge invariant action, and dividing the result by
the volume of the gauge group. The zero modes associated with the translation of the
instanton (anti-instanton) are kept unintegrated till the end. 
\end{enumerate}

We shall now elaborate on the zero mode integration in a bit more detail.
For open strings with both ends on the D-instanton the analysis is the same as for the
D-instanton of type IIB string theory and we can borrow the results of that paper
keeping in mind that we have now similar contribution also from open strings with
both ends on the anti-instanton. We expand the classical open string field with both
ends on the instanton as
\beq \label{state1}
\ksi = i \phi \beta_{-\frac{1}{2}} \, c_0 \, c_1 \ket{-1} 
+ \xi \, c_1 \, d_{-\frac{1}{2}} \ket{-1} +\cdots\, ,
\eeq
where $\ket{-1} $ denote the $-1$ picture vacuum and $d_{-1/2}$ is the fermionic
oscillator associated with the superpartner of the
Euclidean time direction. $\cdots$ denote linear
combination of states with higher $L_0$ eigenvalues.
There are a similar set of modes of the open strings with both ends on the
anti-D-instanton. We shall denote then by $\bar \phi$ and $\bar\xi$ respectively. 

If we consider the open string gauge transformation parameter
$|\theta\rangle =i \theta \beta_{-1/2} c_1|-1\rangle$, then
under the linearized gauge transformation the string field changes by $Q_B|\theta\rangle$
where $Q_B$ is the BRST charge. If the state $\beta_{-1/2} c_1|-1\rangle$ had a non-zero
$L_0$ eigenvalue $h$ then $Q_B|\theta\rangle$ would have a component 
$i h\theta \beta_{-\frac{1}{2}} \, c_0 \, c_1 \ket{-1}$ due to the $c_0L_0$ term in $Q_B$.
Comparison with \eqref{state1} shows that this could be interpreted as a
gauge transformation of $\phi$ by $\phi\to\phi+h\theta$. We could then fix the Siegel
gauge by choosing $\phi=0$, producing a Jacobian factor of $h$. This in turn could
be represented as the integral over Fadeev-Popov ghost field $p$ and $q$ with action
$hpq$. In the language of string field theory, these ghost fields are the coefficients of the
Siegel gauge states $i\gamma_{-\frac{1}{2}} \, c_{1}  \ket{-1}$ and
$i\beta_{-\frac{1}{2}} \,  c_1 \ket{-1}$
 in the expansion of the string field.
 
 In actual practice we have $h=0$ and this is the origin of the pair of ghost zero modes in
 the Siegel gauge. This shows that Siegel gauge is not a good choice of gauge and we must go
 back to the original gauge invariant form of the path integral where we integrate over $\phi$
 and divide by the volume of the gauge  group generated by $\theta$. Since the details are
 identical to those in \cite{Sen:2021tpp}, we shall only quote the result.
The analysis of \cite{Sen:2021tpp} shows that
the gauge invariant form of the path integral over the zero modes takes the form
\beq\label{einter}
{1 \over \int d\theta} \int {d\xi\over \sqrt{2\pi}}  {d\phi} \,  e^{-\phi^2/4}
{1 \over \int d\bar \theta} \int  {d\bar\xi\over \sqrt{2\pi}}  {d\bar\phi} \, e^{-\bar\phi^2/4},
\eeq
where $\bar\theta$ is a parameter similar to $\theta$
for the anti-D-instanton. 
Note that the integration 
measures for $\phi,\bar\phi$ are not accompanied by the $1/\sqrt{2\pi}$ factors since 
these are out of Siegel gauge modes and were not present in the original path
integral representation of \eqref{einsanti} via \eqref{eq1}, \eqref{eq2}. Instead, the
integration measure over $\phi,\bar\phi$ is fixed by demanding that if we had gauge
fixed \eqref{eZ} using Faddeev-Popov procedure, we would have gotten the integration
over the ghost fields with the same integration measure that emerges from the application
of \eqref{eq1}, \eqref{eq2} to \eqref{einsanti}. The absence of $1/\sqrt{2\pi}$ factor in the
$\phi,\bar\phi$ integral can be traced to the absence of such factors in the second 
equation in \eqref{eq2}\cite{Sen:2021tpp}.

We now turn to the rest of the open string modes living on the 
instanton-anti-instanton system. Unlike 
in \cite{Sen:2021tpp}, we do not have any R-sector states in the present system.
We however have a pair of
`tachyons'\footnote{Even though we call them tachyons, for $\Delta x > \pi\sqrt 2$ they
have positive $L_0$ eigenvalues.}
with $L_0$ eigenvalue $-{1\over 2} +\left( {\Delta x\over 2\pi}\right)^2$ from
open strings stretched between the instanton and the anti-instanton, producing the
second term inside the curly bracket in \eqref{einsanti}. We shall denote them
by $(\psi_1\pm i \psi_2)/\sqrt 2$. In this sector there are no subtleties
with gauge fixing and we can
continue to use Siegel gauge.
Besides the states mentioned above, there are infinite towers of higher $L_0$ states in all
sectors, but we see from \eqref{einsanti} that their contributions cancel 
between grassmann even and odd modes and so we shall ignore
them.
Combining \eqref{einter} with the integration over the tachyonic modes on the
open string sector connecting the instanton to the anti-instanton, we get
\ben\label{eZ}
Z &=&  {1 \over \int d\theta} \int {d\xi\over \sqrt{2\pi}}  {d\phi} \,  e^{-\phi^2/4}
{1 \over \int d\bar \theta} \int  {d\bar\xi\over \sqrt{2\pi}}  {d\bar\phi} \, e^{-\bar\phi^2/4}
\nonumber \\ &&\hskip 1in \times \int {d\psi_1\over \sqrt{2\pi}} e^{-{1\over 2} \left\{ -{1\over 2} + \left( {\Delta x\over 2\pi}\right)^2 \right\}
\psi_1^2} \int {d\psi_2\over \sqrt{2\pi}} e^{-{1\over 2} \left\{ -{1\over 2} + \left( {\Delta x\over 2\pi}\right)^2 \right\}
\psi_2^2}\, .
  \een

The analysis of \cite{Sen:2021tpp}
also shows that the variables $\xi$, $\bar\xi$ are related to the shifts in the positions
$x_1,x_2$ of the instanton and the anti-instanton via,
\beq
\delta x_1 = g_o\, \pi\sqrt 2\, \xi, \qquad \delta x_2 = g_o\, \pi\sqrt 2\, \bar\xi\, .
\eeq
Furthermore the gauge transformation parameters $\theta$ is related to the rigid $U(1)$ 
transformation parameters $\tilde \theta$, normalized to have period $2\pi$, via
$\theta=2\tilde\theta/g_o$. This gives $\int d\theta=4\pi/g_o$. A similar result holds for
$\int d\bar\theta$. Substituting these into \eqref{eZ} and carrying out the gaussian
integrals over $\phi$, $\bar\phi$, $\psi_1$, $\psi_2$ we get:
\beq\label{ezfin}
Z= \left( {g_o\over 4\pi}\right)^2 \left({1\over g_o\pi\sqrt 2} \right)^2 {1\over 2\pi}\, 
(2\sqrt{\pi})^2 \int dx_1 dx_2 \, {1\over -{1\over 2} + \left( {\Delta x\over 2\pi}\right)^2 }
={1\over 4\pi^2} \, \int dx_1 dx_2 \, {1\over (\Delta x)^2 - 2\pi^2}\, .
\eeq
Comparing this with \eqref{ezans} we get,\footnote{For this analysis we should regard 
$\NN_D^{2}$ as one unit instead of assigning  separate normalization factors 
to D-instantons and anti-D-instantons, since these suffer from 
infrared divergences from the closed string channel.}
\beq
\NN_D^{2} ={1\over 4\pi^2}\, .
\eeq
This agrees with the value of $\NN_D^{2}$ determined in \cite{Balthazar:2022apu} using comparison with the
matrix model results.

\section{Finite temperature}

For completeness we shall also briefly discuss the finite temperature case, obtained by
compactifying $X^0$ on a circle of radius $R$. In this case the trace over open string
states will include the extra winding modes where the open string wraps the circle 
$\om$ times. From eqs.(3.14) and (3.16)
of \cite{Balthazar:2022apu} we
see that this corresponds to multiplying the zero temperature result \eqref{ezfin} by,
  \beq
I \equiv \exp\left[
\int_0^{\infty} \frac{dt}{t} \sum_{\om \in \ZZZ -\{0\}} \left\{ e^{-2\pi t \lc-{1\over 2}
     + \lc \om R + \frac{\Delta x}{2\pi} \rc^2 \rc} - e^{-2\pi t \lc \om R \rc^2}\right\} \right]\, ,
     \eeq
where we have made the transformations $x\to \sqrt 2 x$, $R\to \sqrt 2 R$ to convert
the results from $\alpha'=2$ unit to $\alpha'=1$ unit. Using \eqref{eq1} we can express
this as,
 \beq
     I= \prod_{\om \in \ZZZ -\{0\}} \frac{(\om R)^2}{\left\{ -{1\over 2}
     + \lc \om R + \frac{\Delta x}{2\pi}\rc^2 \right\}}\, .
  \eeq
 This gives, after minor reorganization of the terms,
  \ben
{I} &=&  
\prod_{\om \in \ZZZ_+} \frac{(\om R)^4}{\lc {1\over \sqrt 2} +\om R - \frac{\Delta x}{2\pi} \rc \lc {1\over \sqrt 2} -\om R - \frac{\Delta x}{2\pi} \rc \lc
{1\over \sqrt 2} +\om R + \frac{\Delta x}{2\pi} \rc \lc 
{1\over \sqrt 2} -\om R + \frac{\Delta x}{2\pi} \rc}\nonumber \\
&=& 
\prod_{\om \in \ZZZ_+} \dfrac{1}{\lc 1 - \dfrac{\lc\frac{\Delta x -\sqrt 
2\pi}{2\pi R}\rc^2}{\om^2} \rc \lc 1 - \dfrac{\lc\frac{\Delta x +\sqrt
2\pi}{2\pi R}\rc^2}{\om^2} \rc}
\, .
  \een
 Using the result,
  \beq
   \sin \theta =  \theta\,  \prod_{\om \in \ZZZ_+} \lc 1 - {\theta^2\over \pi^2 \om^2}\rc \, ,
   \eeq
we can write
\beq
I = {(\Delta x)^2 - 2\pi^2\over 4\, R^2} \, { 1\over \sin \lc \frac{\Delta x + \sqrt 2\pi}{2R}\rc
\sin \lc \frac{\Delta x - \sqrt 2\pi}{2R}\rc}\, .
\eeq
Multiplying \eqref{ezfin} by this factor we get the finite temperature partition function:
\beq
{1\over 4\pi^2} \int dx_1 dx_2 \, {1 \over 4\, R^2\, \sin \lc \frac{\Delta x + \sqrt 2\pi}{2R}\rc
\sin \lc \frac{\Delta x - \sqrt 2\pi}{2R}\rc}\, .
\eeq
This agrees with eq.(3.17) of \cite{Balthazar:2022apu} after changing $x\to x\sqrt 2$,
$R\to R\sqrt 2$.

\bigskip

\noindent{\bf Acknowledgement}: We wish to thank  Bruno Balthazar, Victor Rodriguez and
Xi Yin for useful communications. 
A.S. is supported by ICTS-Infosys Madhava 
Chair Professorship
and the J.~C.~Bose fellowship of the Department of Science and Technology,

\bibliographystyle{JHEP}

\providecommand{\href}[2]{#2}\begingroup\raggedright\endgroup

\end{document}